\documentclass[11pt,usletter]{article}

\usepackage{graphicx}   % need for figures
\usepackage{amssymb,amsmath,mathabx,amsthm}
\usepackage[hyphens]{url}
\usepackage[usenames,dvipsnames,table,svgnames]{xcolor}
\usepackage[breakable, theorems, skins]{tcolorbox}
\usepackage{listings}
\usepackage{xspace}
\usepackage{wrapfig}
\usepackage[inline]{enumitem}
\usepackage{tabularx}
\usepackage[font=small]{caption}
\usepackage{subcaption}
\usepackage[compact]{titlesec}
\usepackage{array}
\usepackage{fullpage}
\usepackage{multirow}
\usepackage{soul}
\usepackage{floatrow}
\usepackage{fancyhdr}
\usepackage[margin=1.0in]{geometry}
\usepackage{hyperref}
\usepackage{textcomp}

\usepackage{mathpazo}
\makeatletter
\renewcommand\paragraph{\@startsection{paragraph}{4}{\z@}%
                                    {0.5ex \@plus1ex \@minus.2ex}%
                                    {-0.5em}%
                                    {\normalfont\normalsize\bfseries}}
    \makeatother

\hypersetup{
    unicode=false,          % non-Latin characters in Acrobat’s 
    colorlinks=true,       % false: boxed links; true: colored links
    linkcolor=black,          % color of internal links (change box color with linkbordercolor)
    citecolor=NavyBlue,        % color of links to bibliography
    filecolor=magenta,      % color of file links
    urlcolor=NavyBlue           % color of external links
}

\tcbset{enhanced}

\newenvironment{mybox}[1][gray!20]{
    \begin{tcolorbox}[   %% Adjust the following parameters at will.
        breakable,
        left=0pt,
        right=0pt,
        top=0pt,
        bottom=0pt,
        colback=#1,
        colframe=#1,
        width=\dimexpr\textwidth\relax,
        boxsep=5pt,
        arc=0pt,outer arc=0pt,    
    ]
}{
    \end{tcolorbox}
}

\newcounter{resq}%[section]

\newcounter{thrust}%[section]

\newtoggle{release}

\togglefalse{release}
\definecolor{diffstart}{named}{Blue}
\definecolor{mark}{named}{Gray}
\definecolor{diffincl}{named}{Green}
\definecolor{diffrem}{named}{Red}

\lstdefinelanguage{diff}{
    basicstyle=\ttfamily\footnotesize,
    morecomment=[f][\color{diffstart}]{@@},
    morecomment=[f][\bfseries\color{mark}]{>},
    morecomment=[f][\color{diffincl}]{+\ },
    morecomment=[f][\color{diffrem}]{-\ },
    morekeywords={def, while, return, if, for, return, int}
}

\lstdefinelanguage{Transformation}{
    basicstyle=\ttfamily\small,
    morekeywords={Map, Where, Update, Insert, Delete, Filter, Match, ConstNode, CNode, Type,
    RefNode, NSeq, Node, Parent}
}

\usepackage{ulem}

\title{Designing Interfaces for Human-Computer Communication: An On-Going Collection of Considerations}
\author{Elena L. Glassman (Harvard)}

\begin{document}

\rfoot{\small{\thepage}}
\setcounter{page}{1}
\noindent

% one-page project summary
\thispagestyle{empty}

%\begin{center}
%\noindent
\section*{Designing Interfaces for Human-Computer Communication: An On-Going Collection of Considerations}
Elena L. Glassman, Harvard University SEAS

\subsection*{Keywords}
frameworks, human-computer communication, interface design, transformative reflection

\subsection*{Abstract}
While we don’t always use words, communicating what we want to an AI is a conversation—with ourselves as well as with it, a recurring loop with optional steps depending on the complexity of the situation and our request. Any given conversation of this type may include: (a) the human forming an intent, (b) the human expressing that intent as a command or utterance, (c) the AI performing one or more rounds of inference on that command to resolve ambiguities and/or requesting clarifications from the human, (d) the AI showing the inferred meaning of the command and/or its execution on current and future situations or data, (e) the human hopefully correctly recognizing whether the AI’s interpretation actually aligns with their intent. In the process, they may (f) update their model of the AI’s capabilities and characteristics, (g) update their model of the situations in which the AI is executing its interpretation of their intent, (h) confirm or refine their intent, and (i) revise their expression of their intent to the AI, where the loop repeats until the human is satisfied. With these critical cognitive and computational steps within this back-and-forth laid out as a framework, it is easier to anticipate where communication can fail, and design algorithms and interfaces that ameliorate those failure points.
%\newpage

\begin{figure}
    \centering
    \includegraphics[width=\linewidth]{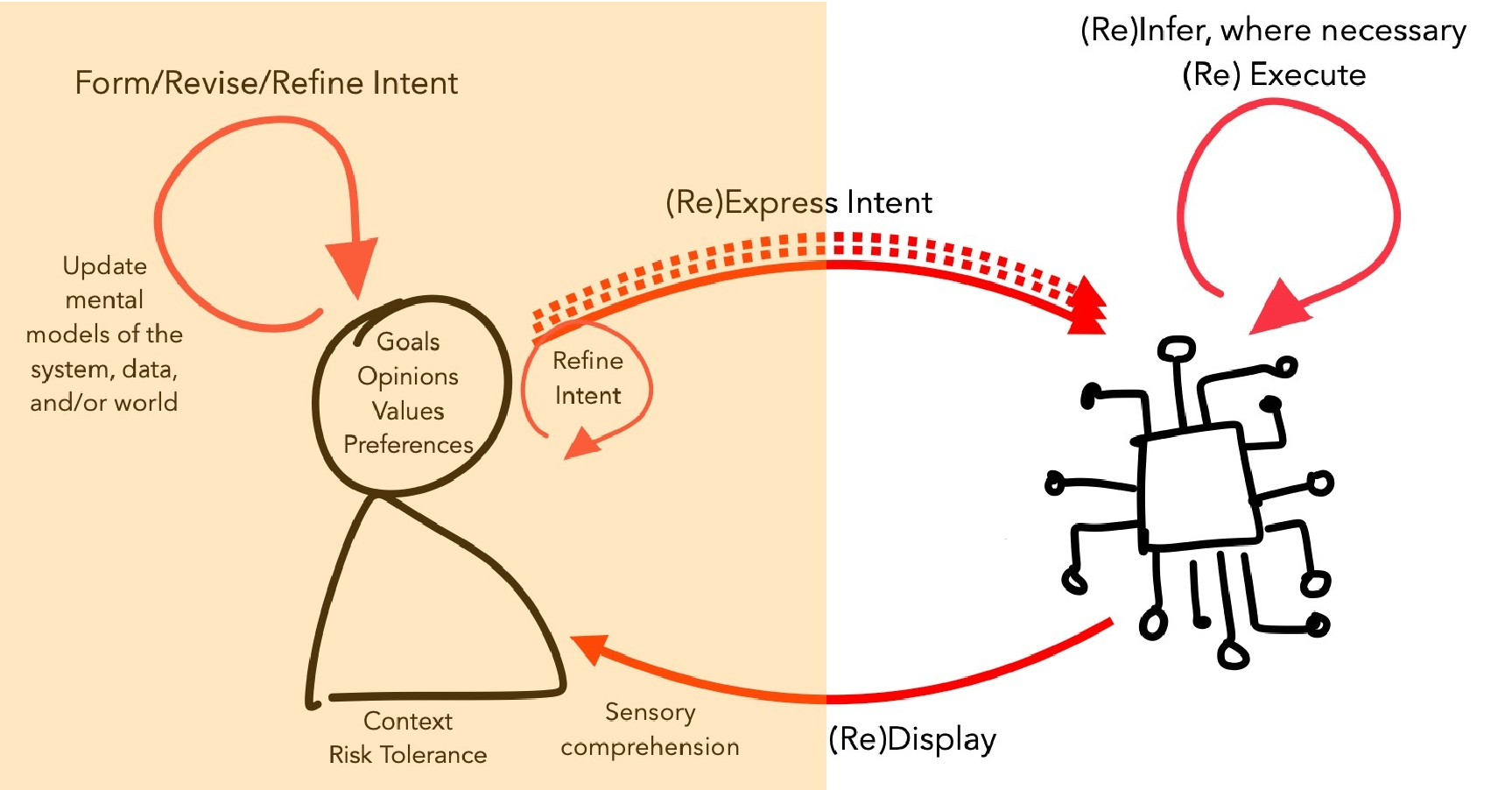}
    \caption{A visual snapshot of the framework in its current form at a high level. The first version was \href{https://www.radcliffe.harvard.edu/event/2021-elena-l-glassman-fellow-presentation-virtual}{presented and recorded at the Radcliffe Institute for Advanced Study} in the Fall of my year in residence (2021-2022). It has evolved as I have done additional systems-building research and incorporated that work back into the framework when presenting at CS venues such as the Berkeley Programming Systems Seminar, Google's PAIR (People + AI Research) Talks, the Intelligent Systems Center Seminar Series at the Johns Hopkins Applied Physics Lab, and the Stanford Seminar on People Computers and Design. Significant feedback was provided by members of the Psychology community at the Cognitive Development Society and APS Annual Convention, especially Prof. Michael C. Frank of Stanford.}
    \label{fig:bigfig}
\end{figure}

\section*{Intent}
In order to communicate something to someone else---or some\textit{thing} else---it is necessary to have something in mind to be communicated, i.e., an intent. It could be
\begin{itemize}
    \item a category (e.g., binary classification), 
    \item a function (e.g., a personal similarity metric), 
    \item a target image in the mind, \item a notion of something to say in natural language (e.g., an email that says no politely), 
    \item an operation (e.g., FlashFill~\cite{gulwani2011automating}), or 
    \item a program.
\end{itemize}
This intent may originate initially from the person, as a natural consequence of their goals, opinions, values, preferences, and context, or it may be a reaction to a system's display, e.g., a representation or example of its capabilities, a dataset, or a view of the world.

\section*{Expression}

The human can express their intent using one or more modalities at the same time, by
\begin{itemize}
    \item composing statements in a language---a natural and/or programming language---including program sketches; 
    \item providing concrete examples, demonstrations, or partially-concrete linguistic sketches; 
    \item annotations of whatever is already in or has just been added to the \textit{common ground} shared by the system and the human (e.g., RAGAE~\cite{zhang2020interactive});
    \item physical gestures, e.g., in open space or on a touch sensitive surface; and
    \item GUI interactions, e.g., physical and virtual button pushes that invoke a particular function.
\end{itemize}
Even for an unchanged intent, the human might even try multiple separate ways to specify what they want in parallel, in case one way of expressing what they want is more effective for the system than another, if the system can work on each specification in parallel threads; this can minimize the human's anxiety about making mistakes in any one specification they provide in more complex domains, e.g., synthesizing programs in a language they do not know~\cite{assuage}.

In the process of expressing their intent, the human may have new insights about their intent that impact their intent and their intent expression even before receiving a response from the system.

Relevant concepts in the historical literature include the Gulf of Execution, referring to when the user struggles to use the affordances given to them to express their intent such that the system correctly interprets them.

\section*{Inference and Execution}
If the user expresses their intent in a way that requires no inference, e.g., as statement(s) in a programming language or as a push of a button that invokes a pre-programmed function, then the system can just execute the expressed intent and reflect any feedback to the user in the next step.

If the intent expression has any semantic ambiguity, it is necessary for the system to incorporate some AI/ML to perform inference about the intended intent. There are multiple types of potential inference errors, such as mistaking one spoken word for another or misinterpreting the semantic meaning of a correctly transcribed natural language utterance.

\section*{Feedback}
Feedback to the user can include:

\begin{itemize}
    \item user-relevant components of the system's state, e.g., how many alarms have been set (see also the "Visibility of System State" recommendation within Nielson's usability heuristics);
    \item any inferences the system made based on the user's intent expression, any model it has of the user, and its own base priors and heuristics;
    \item a view (or preview) of what executing the inferred request does, given real or hypothetical data or situations
\end{itemize}
This feedback becomes part of the common ground shared with the human user and, depending on how it is provided to the user, can be explicitly annotated or edited as part of subsequent rounds of intent expression.

%\section*{Existing Frameworks}
%\todo{summarize Norman's theory here}

\section*{Attention, Comprehension}
While this feedback can be delivered, it may or may not be received by the human due to issues with attention and sensory-level comprehension:
\begin{itemize}
    \item not noticing the information, e.g., visual information being too far away from where they are looking
    \item not noticing the information due to it being encoded in a way they cannot perceive, e.g., due to color-blindness
    \item only delivering information on one channel that is blocked, e.g., audio feedback when a device has been muted
\end{itemize}

If the feedback is received, it still needs to be interpreted and comprehended. For example, the human needs to actively construct the meaning of any feedback delivered in visual, natural, or programming languages.

\section*{Mental Models}
In the process of interpreting feedback, the human may consciously or unconsciously update their mental models of:
\begin{itemize}
    \item themselves
    \item the system
    \item the task(s)
    \item any data at hand, and
    \item the world
\end{itemize} 
in which the system might act in/on now or in the future. Based any pre-existing mental models they had~\cite{norman2014someobservations}, any updates as a result of new information, and the personal factors listed previously, as well as task-specific risk tolerances, the human may refine or entirely revise the intent that drives any subsequent intent expressions.

\section*{Evaluation}
Within this conversational loop, the human has the greatest access to their own goals, values, preferences, and context; as a result, only they can decide when the system has sufficiently understood and can correctly carry out the final version of the intent they have attempted to communicate. 

In additional to traditional measures of usabilty evaluation, system designers can literally count the number of trips around this conversational loop it takes for the human to reach the point of the human confidently and correctly understanding that the system has understood their intent. 

In situations where the system is helping the human make a decision or construct an object, the human may more quickly or better fulfill their goals by authoring the final result themselves in the process of or as a result of interacting with the system, even though the system never correctly understood their intent. Interfaces can explicitly afford this, and when counting conversational loops, this is an alternative place to end. User studies that force the user to keep going, expressing and re-expressing their intent, are misleading.

\section*{Additional Cognition Considerations}
Consuming more information also requires time and cognitive resources that we know are scarce. Our cognition is effortful and limited. Like our natural avoidance of pointless physical exertion~\cite{exercised}, we conserve our mental energy. Our ability to hold and manipulate complex situations and ideas can be expanded in any particular domain through specialized training or augmented with specific tools but will always have its limits. To help us make a decision despite this, we unconsciously deploy heuristics and biases to make \textit{some} judgment, despite even substantial remaining uncertainty~\cite{JudgmentunderUncertainty}, so we can move on with our lives rather than being frozen in indecision. %Arriving at \textit{an answer} or moving forward with \textit{a choice} allows us to release the mental work and resources that got us there, while having no clear resolution requires us to continue retaining the evidence for each of several competing possibilities.
Recent work on cognitive engagement \& incidental learning~\cite{incidentalLearning}, as well as the impacts of cognitive forcing functions~\cite{buccinca2020proxy, buccinca2021trust}, speak to these concerns within human-AI interaction specifically.

\section*{Acknowledgements}
I am grateful for the generous engagement and feedback of the audience members of my past talks on this topic and the 2023 CHI course audience members, as well as both my CS and Psychology colleagues, who have given feedback on these ideas and pointed to relevant materials within and beyond my sub-discipline. This material is also based upon work supported by the National Science Foundation under Grants No. 1955699, 2107391, and 2123965.

%\clearpage
%\pagenumbering{arabic}

%\newpage
\bibliographystyle{abbrv}
\bibliography{refs,kimmerged2019,kimrefactor2019}

\end{document}